\documentclass[reprint,amsmath,amssymb,aps,floatfix]{revtex4-2}
\usepackage{graphicx}
\usepackage{dcolumn}
\usepackage{bm}
\usepackage{hyperref}
\usepackage{lipsum}
\usepackage{subcaption}  
\usepackage{caption}         
\usepackage{float}
\usepackage{tikz}
\usepackage{ragged2e}
\usepackage{placeins}
\usetikzlibrary{positioning, arrows.meta}  
\usepackage{relsize}  
\usepackage{overpic}  
\usepackage{placeins}
\usepackage{amssymb}
\usepackage{amsmath}
\usepackage{algorithm}
\usepackage{algpseudocode}
\usepackage{makecell}
\usepackage{bm}
\usepackage{placeins}
\usepackage[export]{adjustbox}
\usepackage{amsthm}

\theoremstyle{remark}

\usepackage{mathtools}
\usepackage{float}
\usepackage{tikz}
\usepackage{ragged2e}
\usetikzlibrary{positioning, arrows.meta}  
\usepackage{relsize}  
\usepackage{overpic}

\begin{document}

\preprint{APS/123-QED}

\title{Analyzing Parametric Oscillator Ising Machines through the Kuramoto Lens}

\author{Nikhat Khan\textsuperscript{1}, E.M.H.E.B. Ekanayake\textsuperscript{1}, Nicolas Casilli\textsuperscript{2}, Cristian Cassella\textsuperscript{2},\\ Luke Theogarajan\textsuperscript{3}, Nikhil Shukla\textsuperscript{1}}
\affiliation{%
\textsuperscript{1}University of Virginia, Charlottesville, VA, USA,\\\textsuperscript{2}Northeastern University, MA, USA\\\textsuperscript{3}University of California Santa Barbara, CA, USA
}%

\begin{abstract}
Networks of coupled nonlinear oscillators are emerging as powerful physical platforms for implementing Ising machines. Yet the relationship between parametric-oscillator implementations and traditional oscillator-based Ising machines remains underexplored. In this work, we develop a Kuramoto-style, canonical phase description of parametric oscillator Ising machines by starting from the Stuart–Landau oscillator model-- the canonical normal form near a Hopf bifurcation, and a natural reduced description for many parametric oscillator implementations such as the degenerate optical parametric oscillator (DOPO) among others. The resulting phase dynamics combine the usual phase-difference coupling observed in the standard Kuramoto model along with an intrinsic phase sum term that is generated when conjugate coupling is considered. Moreover, our formulation helps explain why explicit second-harmonic driving is unnecessary in parametric oscillators and also reveals how quasi-steady amplitude heterogeneity scales the original strength of the spin interaction with potentially adverse impacts on the solution quality. Our work helps develop a unifying view of the oscillator-based approach to designing Ising machines.
\end{abstract}
                             
\maketitle

\section{Introduction}
Ising machines based on coupled nonlinear oscillators have emerged as a compelling framework for accelerating hard combinatorial optimization problems such as minimizing the Ising Hamiltonian~\cite{mohseni2022ising,todri2024computing}. These systems leverage the inherent parallelism embedded in the dynamics of coupled oscillators that could potentially provide computational advantages such as improving the speed of convergence. Two broad classes of oscillator-based Ising machines are being actively pursued: \textbf{(a)} Parametric oscillator-based Ising machines (PoIM)--- that use parametric oscillators such as degenerate optical parametric oscillators (DOPO) \cite{mcmahon2016fully,wang2013coherent,yamamoto2017coherent,honjo2021100}, Josephson parametric oscillators~\cite{PhysRevB.109.014511}, parametric frequency dividers~\cite{casilli2024parametric} etc.; and \textbf{(b)} Traditional oscillator-based Ising machines (OIMs), realized using oscillator designs such as LC tanks \cite{wang2019new, Wang2021}, relaxation oscillators~\cite{bashar2020experimental, mallick2021overcoming}, spin torque oscillators~\cite{albertsson2021ultrafast}, ring oscillators and their variants~\cite{moy20221,graber2022versatile}, as well as oscillator implementations based on emerging functional materials such as $VO_2$~\cite{maher2024cmos,shukla2016ultra,parihar2017computing} and ovonic threshold switches~\cite{lee2024demonstration}. Although both paradigms for realizing the Ising Hamiltonian have been studied extensively, their theoretical treatments have largely evolved in parallel. Despite the intuitive structural similarities, a unifying formulation that expresses parametric-oscillator-based Ising dynamics in the canonical phase form used for traditional oscillator-based Ising machines remains underdeveloped. Establishing this connection can enable cross-pollination of design strategies and advance the broader field of using dynamical systems for computational applications. 

In this work, we address this gap by deriving a canonical phase-form for parametric-oscillator networks starting from the conjugate Stuart–Landau oscillator pumped at $2\omega$. The motivation behind this choice of oscillator is that parametric oscillator designs such DOPOs, commonly used in Ising machines implementations such as the Coherent Ising Machines~\cite{hamerly2019experimental,haribara2016computational}, can be reduced to this form under near-threshold, single-mode operation. The resulting phase dynamics exhibit both difference-phase (relevant to Kuramoto dynamics) and intrinsic sum-phase couplings, with the latter arising directly when the parametric (conjugate) pathway is considered \cite{han2016amplitude,PhysRevE.108.044207}. We then construct an energy function that is minimized by the above dynamics and show that its fixed points map to the standard Ising energy at binary phases. Our analysis clarifies why explicit second-harmonic driving, often employed in traditional OIM implementations~\cite{Wang2021, Bashar20232}, may be unnecessary in PoIMs. Furthermore, we also demonstrate that quasi-steady amplitude heterogeneity—--a well-documented phenomenon in parametric oscillator-based architectures that degrades the solution quality--—effectively rescales the intrinsic strength of spin interactions. This insight opens additional avenues for managing and potentially exploiting amplitude homogeneity in practical implementations~\cite{inui2022control,leleu2019destabilization,kalinin2022computational}. 

We note that while reductions from Stuart--Landau oscillators to Kuramoto-type phase-form models are well-established in the literature~\cite{dorfler2014synchronization,li2022mean}, our analysis  explores this equivalence in the context of Ising machines. Furthermore, we extend the analysis beyond the conventional linear, frequency-degenerate coupling used in most experimental implementations, by incorporating phase-sensitive conjugate pathways. This broadens the analytical framework available for oscillator-based Ising solvers. 

\section{Stuart-Landau Oscillator-based Ising Machines}
We begin by considering $N$ weakly coupled parametric oscillators pumped at (approximately) $2\omega$. Considering near threshold operation, we approximate their dynamics using the conjugate Stuart-Landau model~\cite{mahmoud2024synchronization} as shown in Eq.~\eqref{eq:slowflow} below:

\begin{equation}
\dot{a}_i
= (\mu_i - \alpha_i |a_i|^2)\,a_i
\;+\; \kappa_i e^{i\phi_p}\,a_i^{\!*}
\;+\; \xi\sum_{j\neq i}\Big( J_{ij}\,a_j \;+\; G_{ij}\,a_j^{\!*}\Big),
\label{eq:slowflow}
\end{equation}

Here, $a_i=r_i e^{i\theta_i}$ is the phasor representing the slowly varying complex amplitude,
$\mu_i$ is the net linear gain–loss term,
$\alpha_i (>0, \text{here})$ is the nonlinear saturation,
$\kappa_i e^{i\phi_p}$ represents the parametric pump at $2\omega$ with phase $\phi_p$, and the couplers provide both \emph{normal} ($J_{ij}$) and  conjugate ($G_{ij}$) pathways. $\xi$ is a scalar whose value is set in order to ensure weak coupling among the oscillators. 

As alluded to earlier, dynamics of parametric oscillators such as DOPOs~\cite{haribara2016computational} (near-threshold, single-transverse-mode and adiabatically eliminated pump) can be reduced to the first two terms on the RHS of Eq.~\eqref{eq:slowflow} (see Appendix \ref{appendix: CIM}). The $\sum_{j}J_{ij}a_j$ term captures the linear inter-oscillator mixing in coupled parametric arrays~\cite{StrinatiPRA2019,haribara2016computational}. The $\sum_{j}G_{ij}a_j^{\!*}$ term represents the conjugate coupling. While conjugate coupling has not typically been considered in practical implementations of DOPO-based Ising machines, it has been considered when modeling phenomena such as a first order transition from an oscillatory state to a death state \cite{PhysRevE.108.044207}.

When the coupling among the oscillators is real i.e., $J_{ij}, G_{ij} \in \mathbb{R}$, and symmetric i.e., $J_{ij}=J_{ji}\;;\; G_{ij}=G_{ji}$, the above dynamics minimize an energy function, which can be defined using Wirtinger calculus as,

\begin{widetext}
\begin{equation}
\begin{aligned}
E(\mathbf a,\mathbf a^{\!*})
=\; \sum_{i=1}^N
\Big[\,
-\mu_i\,|a_i|^2
\;+\;\frac{\alpha_i}{2}\,|a_i|^4
\;-\;\frac{\kappa_i}{2}\,\big(e^{i\phi_p} a_i^{\!*2}+e^{-i\phi_p} a_i^{2}\big)
\Big] -\;\frac{\xi}{2}\sum_{\substack{i,j=1\\ i\neq j}}^N
\Big[\,
J_{ij}\,\big(a_i a_j^{\!*}+a_i^{\!*}a_j\big)
\;+\; G_{ij}\,\big(a_i a_j+a_i^{\!*}a_j^{\!*}\big)
\Big]
\end{aligned}
\label{eq:E}
\end{equation}
\end{widetext}

\noindent While Sun et al., \cite{sun2025enhancing} developed similar formulations, Eq.~\ref{eq:E} represents the full complex form of the energy function with the $2\omega$ pump. To show that $\frac{dE}{dt} \leq 0$, we first compute $-\frac{\partial E}{\partial a_i^{\!*}}$ which can be expressed as, 
\begin{equation}
-\frac{\partial E}{\partial a_i^{\!*}}
=(\mu_i-\alpha_i|a_i|^2)\,a_i
+\kappa_i e^{i\phi_p}\,a_i^{\!*}
+\eta\sum_{j\neq i}\big(J_{ij}\,a_j+G_{ij}\,a_j^{\!*}\big) \label{eq:parE}
\end{equation}

Equation \eqref{eq:parE} proves that $-\frac{\partial E}{\partial a_i^{\!*}}=\dot{a}_i$. Now,

\begin{equation}
\begin{split}
\frac{dE}{dt}
&=\sum_{i=1}^N\!\left(
\frac{\partial E}{\partial a_i}\,\dot a_i
+\frac{\partial E}{\partial a_i^{\!*}}\,\dot a_i^{\!*}
\right)\\
&= -\sum_{i=1}^N\!\left(
\frac{\partial E}{\partial a_i}\,
\frac{\partial E}{\partial a_i^{\!*}}
+ \frac{\partial E}{\partial a_i^{\!*}}\,
\frac{\partial E}{\partial a_i}
\right)\\
&= -\,2\sum_{i=1}^N \left|\frac{\partial E}{\partial a_i^{\!*}}\right|^2
\;\le\; 0,
\label{eq:Edot}
\end{split}
\end{equation}
where we used $\dot a_i=-\frac{\partial E}{\partial a_i^{\!*}}$ and its conjugate.\\
\noindent When $\phi_p=\{0,\pi\}$ and $\theta_i^{'} \in \{0,\pi\}$, the energy function (Eq.~\ref{eq:E}) reduces to,    
\[ 
\begin{split}
&E(\mathbf a,\mathbf a^{\!*})
= \sum_{i=1}^N
\Big[\,
-(\mu_i+\kappa_i)r_i^2
+\frac{\alpha_i}{2}r_i^4
\big)
\Big] \\
-\xi&\sum_{\substack{i,j=1\\ i\neq j}}^N
\Big[\,
J_{ij}r_ir_j\cos(\theta_i-\theta_j)
+G_{ij}r_ir_j\cos(\theta_i+\theta_j)
\Big]
\end{split}
\]  

We note that both the additive phase term $\left(\cos(\theta_i+\theta_j)\right)$ and phase difference term $\left(\cos(\theta_i-\theta_j)\right)$ are equivalent to $s_is_j$ at $\theta_i^{'} \in \{0,\pi\}$. Assuming the spin interactions of the original graph are divided into the $W=\gamma J+(1-\gamma)G$.
\[ 
E(\mathbf a,\mathbf a^{\!*})
= C -\xi\sum_{\substack{i,j=1\\ i\neq j}}^N
Wr_ir_js_is_j
\] 
where $C$ is a constant. If the amplitudes are equal $r_i=r^{*} \quad \forall i=1,2,...,N$ and/or exhibit minimum deviation, then the energy function maps to the Ising Hamiltonian (with an offset) given by,

\[ 
E(\mathbf a,\mathbf a^{\!*})
\approxeq C -\xi r^{\!*2} \sum_{\substack{i,j=1\\ i\neq j}}^N
Ws_is_j
\] 
Additional details on the impact of amplitude heterogeneity have been discussed in the following sections.

\section{Kuramoto-style phase description of parametric-oscillator Ising machines}

To derive the canonical phase form, we recast Eq.~\eqref{eq:slowflow} in amplitude--phase form by writing
$a_i=r_i e^{i\theta_i}$, and separate the real and the imaginary parts. Using the constraints on the coupling (real, symmetric) as well as the pump phase ($\phi_p=0$) described above, we obtain

\begin{widetext}
\begin{equation}
\begin{split}
\dot{r}_i
&=(\mu_i-\alpha_i r_i^2)\,r_i +\kappa_i r_i \cos(2\theta_i)
+\xi\sum_{j\neq i}\!\Big(J_{ij} r_j \cos(\theta_j-\theta_i)
+\;G_{ij} r_j \cos(\theta_i+\theta_j)\Big),
\label{eq:amp-slow-fixed}
\\[8pt]
r_i \dot{\theta}_i
&=-\kappa_i r_i \sin(2\theta_i)
+\xi\sum_{j\neq i}\!\Big(J_{ij} r_j \sin(\theta_j-\theta_i)
-\;G_{ij} r_j \sin(\theta_i+\theta_j)\Big)
\end{split}
\end{equation}
\end{widetext}

We now make the following simplifications:
\begin{enumerate}
    \item \textbf{Above threshold:} $r_i$ quickly approaches a quasi-constant $r^\star$. We will revisit this assumption in the following section.
    \item \textbf{Phase reduction:} Neglect $\dot r_i$ and small $r$-variations, yielding Eq.~\eqref{eq:phase-only-fixed}
    \begin{widetext}
    \begin{equation}
    \begin{split}
    \dot{\theta}_i
    &=-\kappa_i \sin(2\theta_i)
    -\xi \sum_{j \neq i} \Big( J_{ij} \sin(\theta_i - \theta_j)
    + G_{ij} \sin(\theta_i + \theta_j ) \Big)\\
    \label{eq:phase-only-fixed}
    &\equiv -K_s \sin(2\theta_i)
    - K \sum_{j \neq i} \Big( J_{ij} \sin(\theta_i - \theta_j)
    + G_{ij} \sin(\theta_i + \theta_j) \Big)
    \end{split}
    \end{equation}    
    \end{widetext}
     where, $K_s=\kappa_i$ is the equivalent of strength of the second harmonic injection used in OIMs \cite{Wang2021} and we assume that $\kappa_i$ is equal across all the oscillators. $\xi=K$ is the coupling strength (in the weak coupling regime). The change of variables is performed to maintain consistency with the traditional representation of the OIM dynamics.
     \item \textbf{Equal split between normal and conjugate coupling:} When  the normal and conjugate coupling are equally split ($\gamma=0.5$), $J_{ij}=G_{ij}$. Then,    
    \end{enumerate}

\begin{widetext}
    \begin{equation}
        \begin{split}
        \dot{\theta}_i
        &= -K_s \sin(2\theta_i)
        -\frac{K}{2}\sum_{j\neq i} J_{ij}\Big(\sin(\theta_i-\theta_j)+\sin(\theta_i+\theta_j)\Big)\\
        &=-K_s \sin(2\theta_i)-K\sum_{j \neq i} J_{ij} \sin(\theta_i ).\cos(\theta_j)
    \label{eq:sum-diff-fixed}        
        \end{split}
    \end{equation}
\end{widetext}

In Eq.~\eqref{eq:sum-diff-fixed}, the $\sin(\theta_i - \theta_j)$ term resembles the Kuramoto dynamics. The additional $\sin(\theta_i + \theta_j)$ term arises directly from the \emph{conjugate coupling} $G_{ij}$ in Eq.~\eqref{eq:phase-only-fixed}, which interestingly represents the spin-interaction exhibited in the \emph{Dynamical Ising Machine}~\cite{ekanayake2025different}. We note that the equal split condition is not necessary for the dynamics to minimze the Ising Hamiltonian although the dynamical properties such as stability will depend on the exact split.

When $G_{ij}=0$, Eq.~\eqref{eq:sum-diff-fixed} reduces to,
\begin{equation}
        \dot{\theta}_i
        = -K_s \sin(2\theta_i)
        -K\sum_{j\neq i} J_{ij}\sin(\theta_i-\theta_j) \label{eq:OIM}   
\end{equation}   

\noindent represents the canonical OIM (derived using Kuramoto dynamics). In the following paragraphs, we show that Eq.~\eqref{eq:sum-diff-fixed} and its derivative, Eq.~\eqref{eq:OIM}, also minimizes the Ising Hamiltonian.

It can be observed that in coupled parametric oscillators, the second-harmonic component arises intrinsically from the system's nonlinear dynamics. This inherent feature obviates the need for external harmonic injection, as recently demonstrated by Casilli \textit{et al.}~\cite{casilli2024parametric}, who implemented parametric oscillator–based Ising machines without relying on explicit second-harmonic driving.

The energy function minimized by the above dynamics, can be expressed as,
\begin{widetext}
\begin{equation}
\begin{split}
E_{\theta}
&= -\frac{K_s}{2}\sum_{i=1}^N \cos(2\theta_i) 
-\frac{K}{2} \sum_{\substack{i,j=1 \\ j\neq i}}^N
J_{ij}\left( \cos(\theta_i - \theta_j) + \cos(\theta_i + \theta_j) \right) \\
&= -\frac{K_s}{2}\sum_{i=1}^N \cos(2\theta_i)
   - K \sum_{\substack{i,j=1 \\ j\neq i}}^N J_{ij}\,\cos\theta_i\,\cos\theta_j,
\label{eq:energy_phase}
\end{split}
\end{equation}
\end{widetext}

It can be easily shown that $\frac{dE_{\theta}}{dt} \leq 0$ (via $\dot\theta_i=-\frac{\partial E}{\partial\theta_i}$), and at $\theta^{'} \in \{0,\pi\}^N$, the energy function maps to the Ising Hamiltonian:
\begin{equation}
E_{\theta} \equiv -\,2K \sum_{i<j} J_{ij}\, s_i s_j + C_1 \quad
\label{eq:ising-energy}
\end{equation}
\noindent where $s_i=\cos\theta_i\in\{\pm 1\}$ and $C_1=-\frac{NK_s}{2}$ is a constant offset. We also note that the preceding analysis is performed in a rotating frame. In Appendix \ref{appendix:frame}, we discuss how the choice of the frame of reference impacts the dynamics and the observables.

\section{Amplitude Heterogeneity}
We now revisit the assumption on the homogeneity of the amplitudes, $r_i$. When deriving Eq.~\eqref{eq:sum-diff-fixed}, we had assumed that all oscillator amplitudes converge to a common steady-state value \( r^\star \). More generally, however, each oscillator may saturate to a distinct amplitude \( r_i^{\star} > 0 \), introducing heterogeneity into the system. To analyze the implications of this amplitude variation, we begin with the amplitude--phase formulation derived from Eq.~\eqref{eq:amp-slow-fixed}. Prior to amplitude normalization, the phase dynamics, assuming near steady-state amplitude, are governed by,
\begin{equation}
\begin{split}
r_i^{\star} \dot{\theta}_i &= -\kappa_i r_i^{\star} \sin(2\theta_i) \\\\
&\quad -K \sum_{j \neq i} \left(J_{ij}\, r_j^{\star} \sin(\theta_i - \theta_j) + G_{ij}\, r_j^{\star} \sin(\theta_i + \theta_j) \right)
\end{split}
\label{eq:phase-pre-divide}
\end{equation}

Dividing both sides by \(r_i^{\star} > 0 \) and expressing $\kappa_i=K_s$ yields the normalized phase equation:
\begin{equation}
\begin{split}
\dot{\theta}_i &= -K_s \sin(2\theta_i) \\\\
&\quad -K \sum_{j \neq i} \left(J_{ij}\, \frac{r_j^{\star}}{r_i^{\star}} \sin(\theta_i - \theta_j) + G_{ij}\, \frac{r_j^{\star}}{r_i^{\star}} \sin(\theta_i + \theta_j) \right)
\end{split}
\label{eq:phase-post-divide}
\end{equation}

It can be observed that the only modification relative to the homogeneous case Eq.~\eqref{eq:phase-only-fixed} is the appearance of the amplitude ratio \( \frac{r_j^{\star}}{r_i^{\star}} \), which is expected to introduce a form of weighted interaction between oscillators. Importantly, the system retains its gradient flow structure, with the energy function given by,
\begin{equation}
\begin{split}
E_{AH} &= -\frac{K}{4} \sum_{i \neq j} J_{ij} r_i r_j \left[ \cos(\theta_i - \theta_j) + \cos(\theta_i + \theta_j) \right] \\
&\quad - \frac{K_s}{2} \sum_i r_i^2 \cos(2\theta_i)
\end{split}
\end{equation}

\noindent At binary phase states \( \theta_i^{'} \in \{0, \pi\} \), corresponding to spin variables \( s_i = \cos\theta_i \in \{+1,-1\} \), the energy function simplifies to a weighted Ising Hamiltonian with an offset:
\begin{equation}
\begin{split}
E_{AH} &= -\frac{K}{2} \sum_{i < j} (r_i^{\star} r_j^{\star} J_{ij}) s_i s_j - \frac{K_s}{2} \sum_i r_i^{\star2} \\
&\equiv -\frac{K}{2} \sum_{i < j} \tilde{J}_{ij} s_i s_j - C_2
\end{split}
\end{equation}
\noindent where $C_2=\frac{K_s}{2} \sum_i  r_i^{\star2}$ is a constant. This reveals that amplitude heterogeneity effectively rescales the interactions between the oscillators. The modified interactions can be expressed as, \( \boldsymbol{\tilde{J}_{ij} = r_i^{\star} r_j^{\star} J_{ij}} \). Consequently, in the presence of such heterogeneity, the oscillator network effectively minimizes a different Hamiltonian that of the original problem, which is likely to have an adverse impact on the computational performance. Furthermore, the ratio of amplitude heterogeneity can also impact the dynamical properties such as local stability among others~\cite{cheng2024control,ekanayake2025different}.

\begin{figure*}
    \centering
    \includegraphics[width=0.9\textwidth]{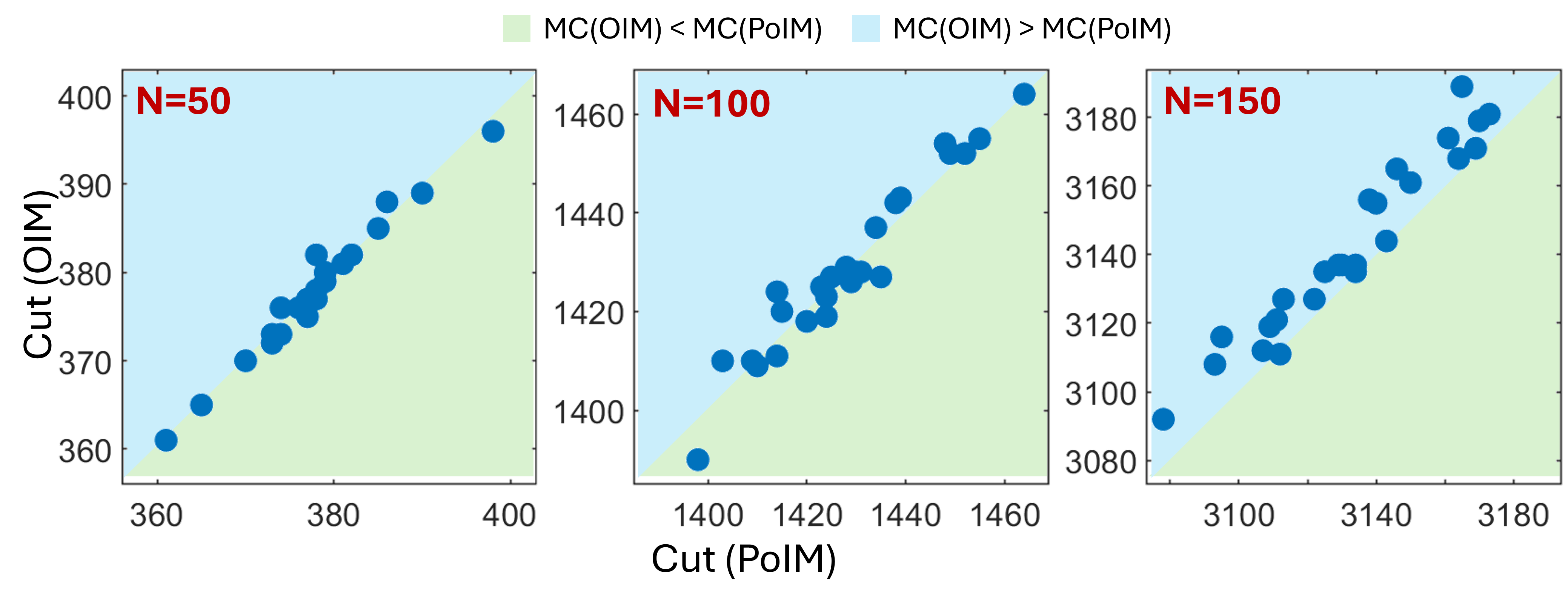}
    \caption{MaxCut solutions obtained using Kuramoto-based OIM dynamics and the canonical phase form of the PoIM dynamics for random generated graphs of varying sizes: (a) $N = 50$; (b) $N = 100$; (c) $N = 150$. For each graph size, 25 independent graph instances are considered, and each instance is simulated 10 times using both models. The best solution from each model is used for comparison.}
    \label{fig:PoIM_OIM}
\end{figure*}

\section{Simulations}
 We test the canonical phase form of the PoIM dynamics (described by Eq.~\ref{eq:sum-diff-fixed}) and the OIM (described by Eq.~\ref{eq:OIM}) on randomly generated graphs of varying sizes. Each graph is generated with an edge probability of \( p = 0.5 \), and for each size \( N = 50, 100, 150 \), we evaluate 25 independent graph instances. For each graph, we compute the MaxCut, which directly corresponds to the solution of the associated anti-ferromagnetic Ising Hamiltonian. Each instance is simulated across 10 independent trials, and the best solution from each set is retained. The simulation parameters used are shown in Appendix~\ref{appendix: simulations}.

Figure~\ref{fig:PoIM_OIM} presents the solutions obtained using the two dynamics. For \( N = 50 \) and \( N = 100 \), the results are closely distributed around the identity line, \( y = x \), indicating no consistent trend favoring one dynamic over the other. For \( N = 150 \), the solutions produced by the OIM are slightly better. However, this observation is not conclusive evidence of superior performance, as both models were simulated using similar parameter settings. It is expected that each model may perform optimally under different conditions, influenced by factors such as stability and convergence behavior. The objective here is to demonstrate that both models faithfully realize the Ising Hamiltonian.

\begin{figure}[htbp!]
    \centering
    \includegraphics[width=0.85\linewidth]{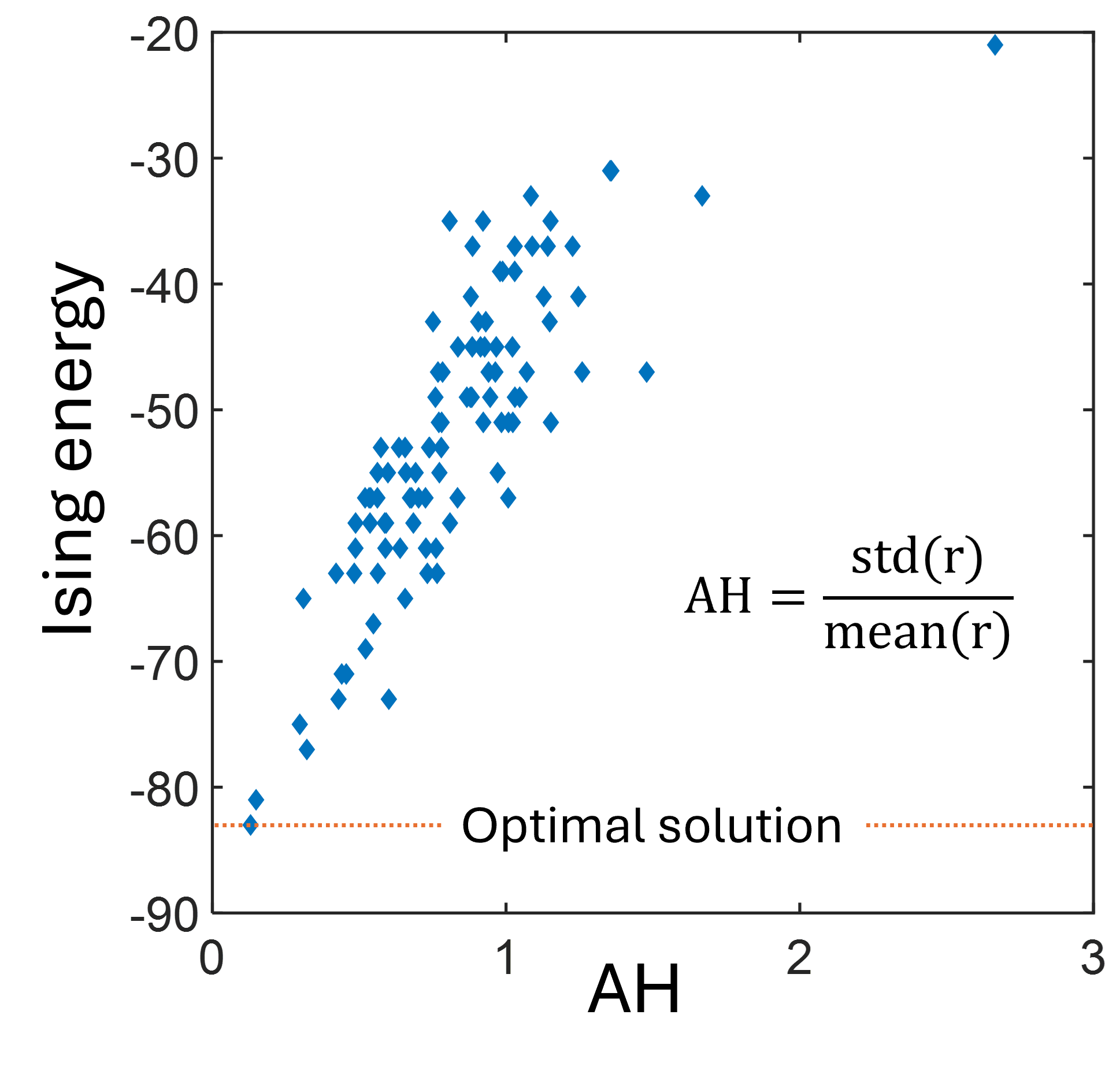}
    \caption{Evolution of the lowest Ising energy obtained across varying levels of amplitude heterogeneity (AH) for an illustrative 50-node regular graph with degree 5. The results corroborate the theoretical result that increasing AH degrades solution quality. Trials=100.}
    \label{fig:AH}
\end{figure}

Figure~\ref{fig:AH} illustrates the impact of amplitude heterogeneity using an illustrative 50-node regular graph with degree 5. To demonstrate this, we evaluate the MaxCut of the graph by simulating the coupled Stuart–Landau oscillators (Eq.~\ref{eq:slowflow}), introducing a small randomness in the gain dispersion parameter (see Appendix \ref{appendix: simulations}). The dynamics are simulated using a stochastic differential equation (SDE) framework over a total time period of $T_{\mathrm{stop}} = 10$. A total of 100 independent trials are performed, with a noise strength of $K_n = 0.08$.

We define the amplitude heterogeneity as the coefficient of variation:
\( \mathrm{AH} = \frac{\mathrm{std}(r)}{\mathrm{mean}(r)} \).
Spins are read out as $s_i = \mathrm{sign}(\cos \theta_i)$, and the achieved Ising energy is evaluated with the effective coupling $W = J + G$ (50/50 split), using \(E = -\frac{1}{2} s^\top W s\). Consistent with theory, it can be observed that the solution quality degrades with increasing amplitude heterogeneity. This also provides a direction for improving solution quality: reduce AH.

\section{Conclusion}
In conclusion, we established the dynamical framework and energy function for a Stuart–Landau oscillator-based Ising machine. Starting from the conjugate coupled Stuart–Landau oscillator model--a natural reduced description for PoIMs, we developed a Kuramoto-style canonical phase form description for PoIMs. Besides the expected equivalence in their capability of realizing the Ising Hamiltonian, our analysis establishes the theoretical underpinnings of important characteristics of PoIMs such as the absence of the need for explicit second harmonic driving and the adversarial impact of amplitude heterogeneity on the computational performance. While our present analysis focuses on the near-threshold, single-mode regime with symmetric, phase-aligned couplers and an adiabatic pump response, future work will relax these constraints to address phase-lagged couplers, non-adiabatic and stochastic amplitude/pump dynamics, heterogeneity in \(p_0\), multi-mode and spatial effects, and conjugate cross-couplings. Furthermore, from a computational standpoint, analyzing the dynamics through the Kuramoto lens to extend the functional properties of PoIMs to higher-order interactions~\cite{bashar2023designing} and to 
$K>2$-state spin systems~\cite{mallick2022computational} represents another promising direction.

\section*{Acknowledgments}
This material was based upon work supported in part by the ARO award W911NF-24-1-0228 and National Science Foundation (NSF) under Grant No. 2328961 and was supported in part by funds from federal agency and industry partners as specified in the Future of Semiconductors (FuSe) program. The work was also supported in part by NSF grant 2435418\\

\noindent \textbf{AUTHOR CONTRIBUTIONS:} \textbf{Nikhat Khan} Validation (lead); Writing – review \& editing (supporting). \textbf{E.M.H.E.B Ekanayake:} Validation (equal); Writing – review \& editing (supporting). \textbf{Nicolas Casilli:} Validation (equal); Writing – review \& editing (supporting). \textbf{Cristian Cassella:} Validation (equal);  Writing – review \& editing (supporting). \textbf {Luke Theogarajan:} Conceptualization (equal) Validation (supporting); Writing – review \& editing (supporting). \textbf {Nikhil Shukla:} Conceptualization (lead); Funding acquisition (lead); Supervision (lead); Validation (equal); Writing – original draft (lead); Writing – review \& editing (lead). \\

\textbf{DATA AVAILABILITY:} The data that support the findings of this study are available from the corresponding author upon reasonable request.

\vspace{1in}

\appendix
\section{Reducing DOPO dynamics to Conjugate Stuart-Landau Model}
\label{appendix: CIM}
As elucidated by Haribara et al.,~\cite{haribara2016computational}, the in-phase and quadrature-phase amplitudes of a single isolated DOPO pulse obey the following c-number stochastic differential equations:
\begin{flushleft}
\textbf{In-phase component:}
\end{flushleft}
\begin{equation}
\dot c=(-1+p_0)c-(c^2+s^2)c \label{eq:in-phase}
\end{equation}

\begin{flushleft}
\textbf{Quadrature component:}
\end{flushleft}
\begin{equation}
\dot s=-(1+p_0)s-(c^2+s^2)s \label{eq:quad}
\end{equation}

In the above equations, the pump field has been adiabatically eliminated by assuming that the pump photon decay rate $\gamma_p$ is much larger than the signal photon decay rate $\gamma_s$. Furthermore, the diffusion terms have also been neglected in this analysis. Equations~\eqref{eq:in-phase} and ~\eqref{eq:quad} can be expressed as,

\begin{equation}
    \begin{split}
        \dot c+i \dot s&=-(c+is)\\\\ &
        - (c^2+s^2)(c+is)+p_0(c-is)
    \end{split}
\end{equation}

Defining $z=c+is$,

\begin{equation}
    \begin{split}
    \dot z=-z -|z|^2z+ p_0z^{\star} \label{eq:complex}
    \end{split}
\end{equation}

To generalize the model to allow for pump injection with a variable phase offset \( \phi_p \), the Eq.~\ref{eq:complex} becomes:
\begin{equation}
    \begin{split}
    \dot z=-z -1.|z|^2z+ p_0 e^{i\phi_p}z^{\star}
    \end{split}
\end{equation}

We rescale the amplitude and the time as, $a=A_sz$, where $A_s^2=\frac{\gamma_s}{\alpha}$, $p_0=\frac{\kappa}{\gamma_s}$ and $\tau=\frac{t}{\gamma_s}$. 
\begin{equation}
    \begin{split}
    \frac{\dot zA_s}{\gamma_s}&=-zA_s -\frac{A_s^2\alpha}{\gamma_s}|z|^2A_sz+ \frac{\kappa}{\gamma_s} e^{i\phi_p}A_sz^{\star} \label{eq:general_form}\\\\
    \Rightarrow \dot a&=-\gamma_sa -\alpha |a|^2a + \kappa e^{i\phi_p}a^{\star}
    \end{split}
\end{equation}

Defining $\mu=-\gamma_s$
\begin{equation}
    \begin{split}
\dot a&=(\mu  -\alpha |a|^2)a + \kappa e^{i\phi_p}a^{\star}
    \end{split}
\end{equation}

This is the Stuart-Landau model for a single oscillator considered in Eq.~\eqref{eq:slowflow}. When weakly coupled with other oscillators, the dynamics of oscillator $i$ can be expressed as,
\begin{equation}
    \dot{a}_i
= (\mu_i - \alpha_i |a_i|^2)\,a_i
\;+\; \kappa_i e^{i\phi_p}\,a_i^{\!*}
\;+\; \xi\sum_{j\neq i}\Big( J_{ij}\,a_j \;+\; G_{ij}\,a_j^{\!*}\Big),
\label{eq:SL}
\end{equation}

\section{Frame of Reference}
\label{appendix:frame}
The canonical phase formulation also helps understand the effect of the frame of reference. As noted earlier, the above analysis was performed in a rotating frame of reference (rotation rate = $\omega$). We also evaluate if these results are valid i.e., if the dynamics minimize the Ising Hamiltonian, in a stationary frame of reference defined by $\theta=\omega t\;+\;\phi \Rightarrow \theta= t+\phi$ (assumption: $\omega=1$). \emph{Here, we clarify that the stationary and the rotating frame nomenclature is defined with respect to the phase variable of the oscillators, and not the observable i.e., $\cos(t+\theta)$}. To begin with, the spin state in the new frame of reference can be defined as $s=\cos(t+\phi)$, and the corresponding dynamics described in Eq.~\eqref{eq:sum-diff-fixed} can then expressed as,
\begin{equation}
\begin{split}
1+\frac{d \phi_i}{dt}&=-\frac{K}{2}\sum_{\substack{j=1\\ j \neq i}} J_{ij} \sin\left(\phi_i-\phi_j\right)\\
&-\frac{K}{2}\sum_{\substack{j=1\\ j \neq i}}J_{ij}\sin\left(2t+\phi_i+\phi_j\right)\\\\
\Rightarrow\frac{d \phi_i}{dt}&=-\frac{K}{2}\sum_{\substack{j=1\\ j \neq i}} J_{ij} \sin\left(\phi_i-\phi_j\right) \\
&-\frac{K}{2}\sum_{\substack{j=1\\ j \neq i}}J_{ij}\sin\left(2t+\phi_i+\phi_j\right)-1
\label{eq:stable_spin_dynamics}
\end{split}
\end{equation}

Interestingly, the steady-state solutions of the above equation correspond to,
\begin{equation}
\begin{split}
\boldsymbol{\phi^{'}}(t) &\in \left\{ -t + 2n\pi,\ -t + (2n+1)\pi \mid n \in \mathbb{Z} \right\}^N\\\\
&\Rightarrow \phi^{'}(t) = -t+ c, \quad \text{where } c \in \{0, \pi\}
\label{eq:fixed_points}
\end{split}
\end{equation}
\noindent Here, $c$ is expressed in the wrapped form. In fact, at steady state $c\equiv\theta^{'} \Rightarrow \phi_i(t)=-t+\theta_i^{'}$. The above dynamics minimize the following energy function 
\begin{equation}
\begin{split}
E^{st} &=-\frac{K}{2}\sum_{\substack{i=1}}^N\left(J_{ij}\cos\left(t+\phi_i\right)\left(\sum_{\substack{j=1}}^N\cos\left(t+\phi_j\right)\right)\right)\\\\
&= -\frac{K}{4}  \sum_{i,j; j\neq i} J_{ij}\left( \cos(\phi_i - \phi_j) + \cos(2t+\phi_i + \phi_j) \right)
\end{split}
\end{equation}
\noindent $E^{st}$ is obtained by substituting $\theta=t+\phi$. To show $\frac{dE^{st}}{dt}\leq0 $, we express,

\begin{equation}
\frac{dE^{st}}{dt}=\sum_{\substack{i=1}}^N{\frac{\partial E^{st}}{\partial \phi_i}}.\frac{d\phi_i}{dt} + \frac{\partial E^{st}}{\partial t}  \label{eq:energy_rate} 
\end{equation}

\noindent where, \(\frac{\partial E^{st}}{\partial t}\) can be can be expressed as,
\begin{equation}
\begin{split}
\quad \frac{\partial E^{st}}{\partial t} &=\frac{K}{2}\sum_{\substack{i=1}}^N\sum_{\substack{j=1}}^NJ_{ij}\big(\sin\left(t+\phi_i\right).\cos\left(t+\phi_j\right)\\\\
&\quad +\sin\left(t+\phi_j\right).\cos\left(t+\phi_i\right)\big)\\\\ \label{eq:partial_energy_time}
    &=\frac{K}{2}\sum_{\substack{i=1}}^N\sum_{\substack{j=1}}^N2\left(J_{ij}\left(\sin\left(t+\phi_i\right).\cos\left(t+\phi_j\right)\right)\right) \\
        &=K\sum_{\substack{i=1}}^N\sum_{\substack{j=1}}^NJ_{ij}\left(\sin\left(t+\phi_i\right).\cos\left(t+\phi_j\right)\right)
\end{split}
\end{equation}

\noindent Next, we calculate,
\begin{equation}
\frac{\partial E^{st}}{\partial \phi_i}=K\sum_{\substack{j=1\\ j \neq i}} J_{ij} \sin\left(t+\phi_i\right).\cos\left(t+\phi_j\right)\\\\ \label{eq:partial_energy_phase}
\end{equation}

\noindent From Eq.~\eqref{eq:partial_energy_time} and Eq.~\eqref{eq:partial_energy_phase}, it can be observed that,
\[\frac{\partial E^{st}}{\partial t}=\sum_{\substack{i=1}}^N\frac{\partial E^{st}}{\partial \phi_i}\]

\noindent Substituting the expression for \(\frac{\partial E}{\partial t}\) into Eq.~\eqref{eq:energy_rate},

\begin{align}
\begin{split}
\quad \frac{dE^{st}}{dt}&=\sum_{\substack{i=1}}^N{\frac{\partial E^{st}}{\partial \phi_i}}.\frac{d\phi_i}{dt} + \sum_{\substack{i=1}}^N\frac{\partial E^{st}}{\partial \phi_i} \\\\
&=\sum_{\substack{i=1}}^N{\frac{\partial E^{st}}{\partial \phi_i}.\left(1+\frac{d \phi_i}{d t}\right)}
=\sum_{\substack{i=1}}^N{\frac{\partial E^{st}}{\partial \phi_i}.\left(1+\frac{d \phi_i}{d t}\right)}\\\\
&=-\sum_{\substack{i=1}}^N\left(\frac{\partial E^{st}}{\partial \phi_i}\right)^2 \leq0
\label{eq:decreasing_energy}
\end{split}
\end{align}

\noindent Under quasi-steady state, $\phi(t)=\phi^{'}(t)$, the resulting energy function can be written as,
\begin{equation}
\begin{split}
E^{st} &=-\frac{K}{2}\sum_{i=1}^N \sum_{j=1}^N 
      J_{ij}\cos\!\big(t+\phi_i^{'}\big)\cos\!\big(t+\phi_j^{'}\big) \\[4pt]
   &=-\frac{K}{2}\sum_{i=1}^N \sum_{\substack{j=1 \\ j\neq i}}^N 
      J_{ij}\cos(\theta_i^{'})\cos(\theta_j^{'}) \\[4pt]
   &\equiv K\left(-\tfrac{1}{2}\sum_{i=1}^N\sum_{j=1}^N J_{ij}\,s_i s_j\right)
\end{split}
\label{eq:ising_eq}
\end{equation}
Thus, the energy function in the stationary frame also reduces to the Ising Hamiltonian.
\begin{figure}[h]
    \centering
    \includegraphics[width=1\linewidth]{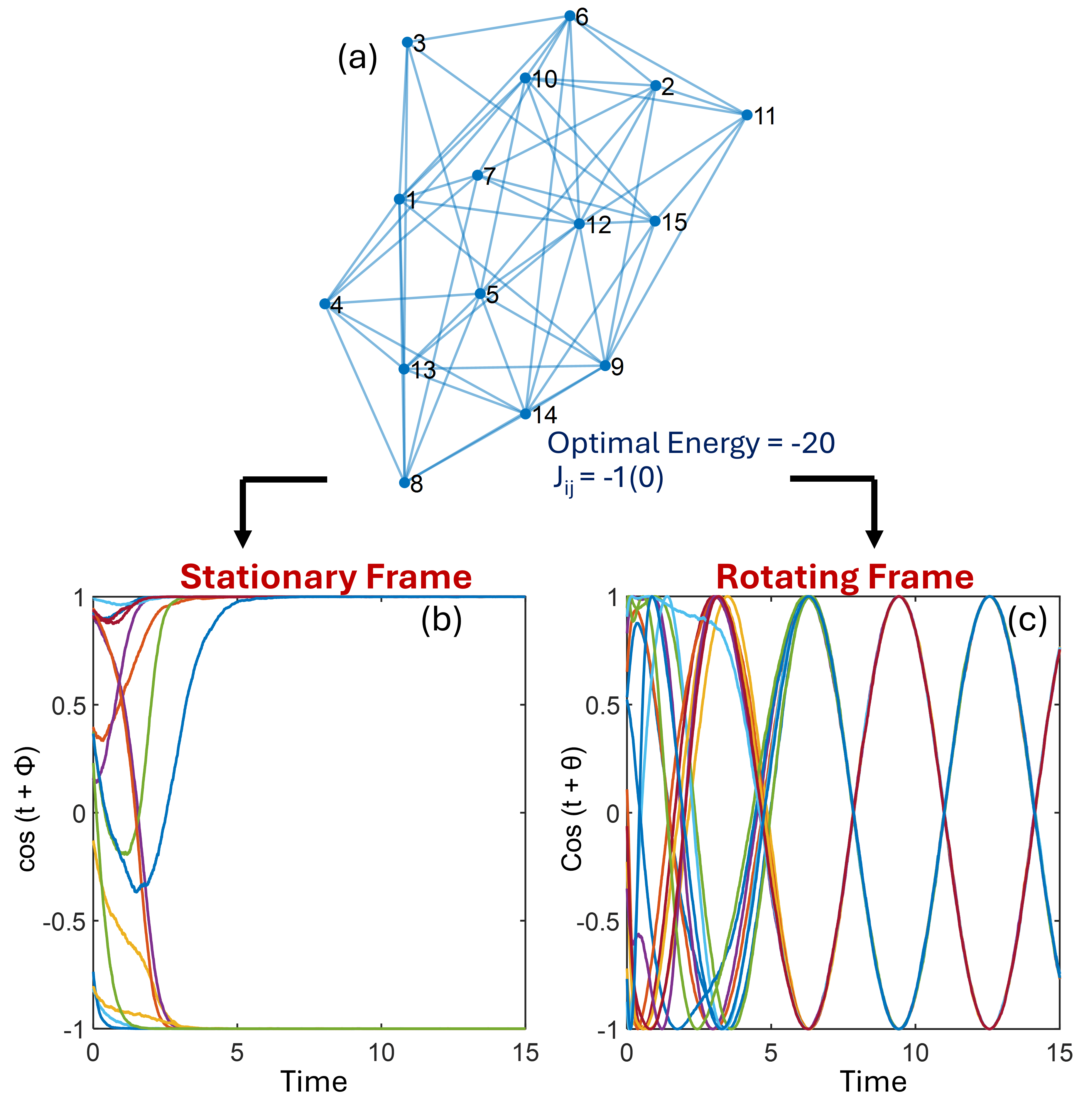}
    \caption{(a) Illustrative graph with 15 nodes and 58 edges. Evolution of the observable output in (b) Stationary; and (c) Rotating frames of reference.}
    \label{fig:time_domain}
\end{figure}

Eq.~\eqref{eq:ising_eq} shows that both the
rotating- and stationary-frame descriptions descend the Ising energy
function. The difference lies only in the observables: in the rotating
frame the spins appear as oscillatory carriers with fixed phase offsets,
while in the lab frame the readout variables directly take static values $\pm 1$. Hence, the two frames are dynamically equivalent in terms of energy minimization, but provide complementary physical perspectives on the same Ising machine.

The above analysis emphasizes a subtle but important distinction: in the rotating frame, the phase remains fixed while the output oscillates; in contrast, in the stationary frame, the dynamics allow the phase to rotate in such a way that the output remains constant (Table~\ref{table:frame}). This also implies that the dynamics described by Eq.~\eqref{eq:stable_spin_dynamics} and Eq.~\eqref{eq:sum-diff-fixed} are equivalent representations of the same system, viewed from different frames of reference.

\begin{table}[h]
\centering
\begin{tabular}{|c|c|c|}
\hline 
\textbf{} & \textbf{Rotating frame} & \textbf{Stationary frame} \\
\hline
\textbf{Phase} & $\theta$ & $\phi(t)= -t+\theta$ \\
\hline
\textbf{Spin state} & $\cos(\theta^{'})$ & $\cos(\theta^{'})$ \\
\hline
\textbf{\shortstack{Osc. output \\ (observable)}} 
& \raisebox{0.5ex}{\shortstack{$\cos(t+\theta)$ \\ (oscillates)}} 
& \raisebox{0.5ex}{\shortstack{$\cos(t+\phi)\Rightarrow\cos(\theta^{'})$ \\ (under steady state)}} \\
\hline
\multicolumn{3}{|c|}{At steady state $\theta^{'} \in \{0,\pi\}$} \\
\hline
\end{tabular}
\caption{Comparison of rotating and stationary frames}
\label{table:frame}
\end{table}

\textit{Simulation}: To demonstrate the differences in the behavior of observables across reference frames, we simulate the representative graph shown in Fig.~\ref{fig:time_domain}(a). Figures~\ref{fig:time_domain}(b) and (c) show the output signal, $\cos(t+\theta)$, computed in the stationary and rotating frames, respectively. In line with the theoretical description described above, in the stationary frame, the observable settles to discrete values in the set $\{-1, +1\}$. In contrast, the rotating frame yields a continuously oscillating output. These distinct behaviors underscore the influence of the chosen reference frame on the observed dynamics.

\section{Simulation Parameters}
\label{appendix: simulations}
\noindent \textbf{1. Simulations to compare OIM with phase form of the PoIM}\\

\noindent \textbf{Traditional OIM Model}
\begin{equation}
\begin{split}
{d}\phi_i
&= \Bigl[
   -K_{OIM} \sum_{\substack{j=1\\j\neq i}}^{N} J_{ij}\,\sin\bigl(\phi_i-\phi_j\bigr) 
   -A_s\,\sin\bigl(2\phi_i\bigr)
  \Bigr]\,{d}t\\\\ 
  & +\; A_n\,{d}W_t \label{eq:SDE1}
\end{split}
\end{equation}

\noindent In Eq.~\ref{eq:SDE1}, $A_s$ represents the strength of the second harmonic injection. The strength of $A_s$ is modulated as, \(A_s=0.5+A_{s,max}.\frac{t}{t_{stop}}\). $K_{OIM}$ is the coupling strength among the oscillators. $A_n$ is the strength of the Gaussian white noise. \\

\noindent \textbf{Phase form of PoIM}
\begin{equation}
\begin{split}
{d}\phi_i
&= \Bigl[
   -K_{PoIM} \sum_{\substack{j=1\\j\neq i}}^{N} J_{ij}\,\sin\bigl(\phi_i-\phi_j\bigr)+\sin\bigl(\phi_i+\phi_j\bigr)     
  \Bigr]{d}t \\\\
  &+A_n{d}W_t \label{eq:SDE2}
\end{split}
\end{equation}
In Eq.~\ref{eq:SDE2}, $K_{PoIM}$ the coupling strength among the oscillators. The strength of $K_{PoIM}$ is modulated as, \(K_{PoIM}=0.5+K_{PoIM,max}.\frac{t}{t_{stop}}\). Furthermore, we assume that $J_{ij}=G_{ij}$. The parameters used in the simulations are shown in Table~\ref{table:PoIM_OIM}.

\begin{table}[!h]
\centering
\begin{tabular}{|c|c|c|c|}
\hline
{} & \textbf{N=50} & \textbf{N=100} & \textbf{N=150} \\
\hline
$K_{OIM}$ (OIM) & 1 & 1 & 1 \\
\hline
$A_{s,max}$ (OIM) & 2 & 3 & 4 \\
\hline
$K_{PoIM,max}$ (PoIM) & 2 & 3 & 4 \\
\hline
$t_{stop}$ (PoIM and OIM) & 10 & 10 & 15 \\
\hline
$A_n$ (PoIM and OIM)& 0.05 & 0.08 & 0.15 \\
\hline
\end{tabular}
\caption{Simulation parameters for OIM and PoIM (phase form) in Fig. 1}
\label{table:PoIM_OIM}
\end{table}

\FloatBarrier

\noindent \textbf{2. Simulations to investigate the impact of Amplitude Heterogeneity}
\begin{equation}
\begin{split}
\dot{r}_i 
&= (\mu_i - \alpha_i r_i^2) r_i + \kappa_i r_i \cos(2\theta_i)\\\\ 
&+ \sum_{j=1}^{N} \left[ J_{ij} \cos(\theta_i - \theta_j) r_j + G_{ij} \cos(\theta_i + \theta_j) r_j \right] \\\\
\dot{\theta}_i 
&=  -\kappa_i  \sin(2\theta_i) \\\\
&- \frac{1}{r_i} \sum_{j=1}^{N} \left( J_{ij} \sin(\theta_i - \theta_j) r_j + G_{ij} \sin(\theta_i + \theta_j) r_j \right)     
\end{split}
\end{equation}

\noindent A small randomness is introduced  the gain dispersion parameter $\mu_i$ to amplify the amplitude heterogeneity. The parameters used in the simulations are shown in Table~\ref{table:AH}.

\begin{table}[h!]
\centering
\begin{tabular}{|c|c|}
\hline
\textbf{Parameter} & \textbf{Value} \\
\hline
$\mu_i$ & $0.6(1+0.2\;\mathcal{N}(0, 1))$ \\
\hline
$\kappa$ & 0.05 \\
\hline
$\alpha$ & 1 \\
\hline
$A_n$ & 0.05 \\
\hline
\end{tabular}
\caption{Parameters used to simulate amplitude heterogeneity in Fig. 2}
\label{table:AH}
\end{table}

\def\bibsection{\section*{References}}  
\bibliography{Stable_spin4_copy}

\begin{thebibliography}{36}

\bibitem{mohseni2022ising}
N. Mohseni, P. L. McMahon, and T. Byrnes, Ising ma
chines as hardware solvers of combinatorial optimization problems, Nature Reviews Physics 4, 363 (2022).

\bibitem{todri2024computing}
A. Todri-Sanial, C. Delacour, M. Abernot, and F. Sabo, Computing with oscillators from theoretical underpin
nings to applications and demonstrators, Npj unconven
tional computing 1, 14 (2024).

\bibitem{mcmahon2016fully}
P. L. McMahon, A. Marandi, Y. Haribara, R. Hamerly, C. Langrock, S. Tamate, T. Inagaki, H. Takesue, S. Ut
sunomiya, K. Aihara, et al., A fully programmable 100
spin coherent Ising machine with all-to-all connections, Science 354, 614 (2016).

\bibitem{wang2013coherent}
Z. Wang, A. Marandi, K. Wen, R. L. Byer, and Y. Ya
mamoto, Coherent Ising machine based on degenerate optical parametric oscillators, Physical Review A 88, 063853 (2013).

\bibitem{yamamoto2017coherent}
Y. Yamamoto, K. Aihara, T. Leleu, K.-i. Kawarabayashi, S. Kako, M. Fejer, K. Inoue, and H. Takesue, Coherent Ising machines—optical neural networks operating at the quantum limit, npj Quantum Information 3, 49 (2017).

\bibitem{honjo2021100}
T. Honjo, T. Sonobe, K. Inaba, T. Inagaki, T. Ikuta, Y. Yamada, T. Kazama, K. Enbutsu, T. Umeki, R. Kasa
hara, et al., 100,000-spin coherent Ising machine, Science advances 7, eabh0952 (2021).

\bibitem{PhysRevB.109.014511}
S. Razmkhah, M. Kamal, N. Yoshikawa, and M. Pe
dram, Josephson parametric oscillator based Ising ma
chine, Phys. Rev. B 109, 014511 (2024).

\bibitem{casilli2024parametric}
N. Casilli, T. Kaisar, L. Colombo, S. Ghosh, P. X.
L. Feng, and C. Cassella, Parametric frequency divider based Ising machines, Physical review letters 132, 147301 (2024).

\bibitem{wang2019new}
T. Wang, L. Wu, and J. Roychowdhury, New compu
tational results and hardware prototypes for oscillator
based Ising machines, in Proceedings of the 56th Annual Design Automation Conference 2019 (2019) pp. 1–2.

\bibitem{Wang2021}
T. Wang, L. Wu, P. Nobel, and J. Roychowdhury, Solv
ing combinatorial optimisation problems using oscillator based Ising machines, Natural Computing 20, 287 (2021).

\bibitem{bashar2020experimental}
M. K. Bashar, A. Mallick, D. S. Truesdell, B. H. Calhoun, S. Joshi, and N. Shukla, Experimental demonstration of a reconfigurable coupled oscillator platform to solve the max-cut problem, IEEE Journal on Exploratory Solid
State Computational Devices and Circuits 6, 116 (2020).

\bibitem{mallick2021overcoming}
A. Mallick, M. Bashar, D. Truesdell, B. Calhoun, and N. Shukla, Overcoming the accuracy vs. performance trade-off in oscillator Ising machines, in 2021 IEEE International Electron Devices Meeting (IEDM) (IEEE, 2021) pp. 40–2.

\bibitem{albertsson2021ultrafast}
D. I. Albertsson, M. Zahedinejad, A. Houshang, R. Khymyn, J. \AA{}kerman, and A. Rusu, Ultrafast Ising machines using spin torque nano-oscillators, Applied Physics Letters 118 (2021).

\bibitem{moy20221}
W. Moy, I. Ahmed, P.-w. Chiu, J. Moy, S. S. Sapatnekar, and C. H. Kim, A 1,968-node coupled ring oscillator cir
cuit for combinatorial optimization problem solving, Na
ture Electronics 5, 310 (2022).

\bibitem{graber2022versatile}
M. Graber and K. Hofmann, A versatile \& adjustable 400 node cmos oscillator based Ising machine to inves
tigate and optimize the internal computing principle, in 2022 IEEE 35th International System-on-Chip Confer
ence (SOCC) (IEEE, 2022) pp. 1–6.

\bibitem{maher2024cmos}
O. Maher, M. Jim´enez, C. Delacour, N. Harnack, J. N´u˜nez, M. J. Avedillo, B. Linares-Barranco, A. Todri
Sanial, G. Indiveri, and S. Karg, A cmos-compatible oscillation-based vo2 Ising machine solver, Nature Com
munications 15, 3334 (2024).

\bibitem{shukla2016ultra}
N. Shukla, W.-Y. Tsai, M. Jerry, M. Barth, V. Narayanan, and S. Datta, Ultra low power coupled os
cillator arrays for computer vision applications, in 2016 IEEE symposium on VLSI technology (IEEE, 2016) pp. 1–2.

\bibitem{parihar2017computing}
A. Parihar, N. Shukla, M. Jerry, S. Datta, and A. Ray
chowdhury, Computing with dynamical systems based on insulator-metal-transition oscillators, Nanophotonics 6, 601 (2017).

\bibitem{lee2024demonstration}
Y. W. Lee, S. J. Kim, J. Kim, S. Kim, J. Park, Y. Jeong, G. W. Hwang, S. Park, B. H. Park, and S. Lee, Demon
stration of an energy-efficient Ising solver composed of ovonic threshold switch (ots)-based nano-oscillators (ot
snos), Nano Convergence 11, 20 (2024).

\bibitem{hamerly2019experimental}
R. Hamerly, T. Inagaki, P. L. McMahon, D. Venturelli, A. Marandi, T. Onodera, E. Ng, C. Langrock, K. Inaba, T. Honjo, et al., Experimental investigation of perfor
mance differences between coherent Ising machines and a quantum annealer, Science advances 5, eaau0823 (2019).

\bibitem{haribara2016computational}
Y. Haribara, S. Utsunomiya, and Y. Yamamoto, Compu
tational principle and performance evaluation of coherent Ising machine based on degenerate optical parametric os
cillator network, Entropy 18, 151 (2016).

\bibitem{han2016amplitude}
W. Han, H. Cheng, Q. Dai, H. Li, P. Ju, and J. Yang, Amplitude death, oscillation death, wave, and multista
bility in identical stuart–landau oscillators with conju
gate coupling, Communications in Nonlinear Science and Numerical Simulation 39, 73 (2016).

\bibitem{PhysRevE.108.044207}
R. Ghosh, U. K. Verma, S. Jalan, and M. D. Shrimali, First-order transition to oscillation death in coupled os
cillators with higher-order interactions, Phys. Rev. E 108, 044207 (2023).

\bibitem{Bashar20232}
M. K. Bashar, Z. Lin, and N. Shukla, Stability of oscil
lator Ising machines: Not all solutions are created equal, J. Appl. Phys. 134, 144901 (2023).

\bibitem{inui2022control}
Y. Inui, M. D. S. H. Gunathilaka, S. Kako, T. Aonishi, and Y. Yamamoto, Control of amplitude homogeneity in coherent Ising machines with artificial zeeman terms, Communications Physics 5, 154 (2022).

\bibitem{leleu2019destabilization}
T. Leleu, Y. Yamamoto, P. L. McMahon, and K. Aihara, Destabilization of local minima in analog spin systems by correction of amplitude heterogeneity, Physical review letters 122, 040607 (2019).

\bibitem{kalinin2022computational}
K. P. Kalinin and N. G. Berloff, Computational complex
ity continuum within Ising formulation of np problems, Communications Physics 5, 20 (2022).

\bibitem{dorfler2014synchronization}
F. D¨orfler and F. Bullo, Synchronization in complex net
works of phase oscillators: A survey, Automatica 50, 1539 (2014).

\bibitem{li2022mean}
Y. Li, J. Shi, and K. Aihara, Mean-field analysis of stuart–landau oscillator networks with symmetric cou
pling and dynamical noise, Chaos: An Interdisciplinary Journal of Nonlinear Science 32 (2022).

\bibitem{mahmoud2024synchronization}
M. Mahmoud, M.Medhat,andH.F.El-Nashar,Synchro
nization of two indirectly coupled singly resonant optical parametric oscillators, Optical and Quantum Electronics
56, 956 (2024). 11.

\bibitem{StrinatiPRA2019}
M. C. Strinati, L. Bello, A. Pe'er, and E. G. D. Torre, Theory of coupled parametric oscillators beyond coupled Ising spins, Physical Review A 100, 023835 (2019).

\bibitem{sun2025enhancing}
L. Sun, M. X. Burns, and M. C. Huang, Enhancing oscillator-based Ising machine models with amplitude dynamics and polynomial interactions, arXiv preprint arXiv:2504.00329 (2025).

\bibitem{ekanayake2025different}
E. Ekanayake and N. Shukla, Different paths, same des
tination: Designing physics-inspired dynamical systems with engineered stability to minimize the Ising hamilto
nian, Physical Review Applied 24, 024008 (2025).

\bibitem{cheng2024control}
Y. Cheng, M. Khairul Bashar, N. Shukla, and Z. Lin, A control theoretic analysis of oscillator Ising machines, Chaos: An Interdisciplinary Journal of Nonlinear Science 34 (2024).

\bibitem{bashar2023designing}
M. K. Bashar and N. Shukla, Designing Ising machines with higher order spin interactions and their application in solving combinatorial optimization, Scientific Reports 13, 9558 (2023).

\bibitem{mallick2022computational}
A. Mallick, M. K. Bashar, Z. Lin, and N. Shukla, Com
putational models based on synchronized oscillators for solving combinatorial optimization problems, Physical Review Applied 17, 064064 (2022).

\end{thebibliography}
\end{document}